\documentclass[12pt]{amsart}

\usepackage[margin=0.9in]{geometry}
\usepackage{graphicx}
\usepackage{subfig}
\usepackage{caption}
\usepackage{xcolor}
\usepackage{cite}
\DeclareMathOperator{\Var}{Var}

\begin{document}

\title{Autoregressive Modeling of Forest Dynamics}

\author{Olga Rumyantseva, Andrey Sarantsev, Nikolay Strigul}

\address{Washington State University in Vancouver, Department of Mathematical Sciences}

\email{olga.rumyantseva@wsu.edu}

\address{University of Nevada in Reno, Department of Mathematics and Statistics}

\email{asarantsev@unr.edu}

\address{Washington State University in Vancouver, Department of Mathematical Sciences}

\email{nick.strigul@wsu.edu}

\begin{abstract}
In this work, we employ autoregressive models developed in financial engineering for modeling of forest dynamics. Autoregressive models have some theoretical advantage over currently employed forest modeling approaches such as Markov chains and individual-based models, as autoregressive models are both analytically tractable and operate with continuous state space. We perform time series statistical analysis of forest biomass and basal area recorded in Quebec provincial forest inventories in 1970--2007. The geometric random walk model adequately describes the yearly average dynamics. For individual patches, we fit an AR(1) process capable to model negative feedback (mean-reversion). Overall, the best fit also turns out to be geometric random walk, however, the normality tests for residuals fail. In contrast, yearly means are adequately described by normal fluctuations, with annual growth, on average, 2.3\%, but with standard deviation of order 40\%. We use Bayesian analysis to account for uneven number of observations per year. This work demonstrates that autoregressive models represent a valuable tool for modeling of forest dynamics. In particular, they  quantify stochastic effects of environmental disturbances and develop predictive empirical models on short and intermediate temporal scales.
\end{abstract}



\maketitle

\thispagestyle{empty}

\section{Introduction}

\subsection{Background}

Understanding the dynamics and self-organization of ecosystems is one of the most challenging problems in modern ecology \cite{Levin1999}. Self-organization occurs simultaneously on several levels of hierarchical ecosystem organization and involves dynamic processes operating on different temporal and spatial scales \cite{Levin2003}. Forest dynamics refers to temporal and spatial changes that occur simultaneously at different levels of ecosystem organization. Various modeling approaches employed to understand and predict these changes include a number of discrete and continuous stochastic and deterministic models, such as Markov chains, individual-based models, ordinary and partial differential equations \cite{Strigul2012}. A large number of forest models have been developed over the last decades \cite{Botkin1993, Shugart1984, pacala1993, pastor2005application, moorcroft2001method, Strigul2008}. Still, there are fundamental questions that existing quantitative approaches fail to fully address. One of the major challenges is the understanding of forest succession and biomass dynamics under the non-stationary disturbance regimes related to climatic factors and anthropogenic activities. An incomplete list of disturbances that substantially affect tree survival and lead to forest biomass decrease include wind, frost, hurricanes, harvesting and forest fires. Markov chains are able to capture effects of all these disturbances on forest biomass dynamics \cite{Lienard16Markov}. However, their application is based on the discretization of the state space \cite{Strigul.et.all.2012}. Spatially-explicit individual-based models are able to simulate effects of disturbances in continuous time and state space \cite{pacala1996, Strigul2008, Strigul2012}. However, these models are typically analytically intractable, i.e. the model predictions are produced by computer simulations only, and it limits our ability to understand what model prediction in general \cite{Strigul2008, Lienard16IBM}.

Another challenge to our understanding of forest dynamics is the multidimensional nature of this complex adaptive system \cite{Levin2003, Lienard2014Quebec}.  Forest disturbances are traditionally associated with a loss of biomass. However, Markov chain models based only on biomass do not capture forest succession comprehensively \cite{Strigul.et.all.2012, Lienard2014Quebec}. This limitation motivates the need for alternative formulations that are able to consider several forest dimensions instead of only one. In our previous study, we combined multivariate statistical analysis with Markov chain approach to develop a multidimensional Markov Chain \cite{Lienard2014Quebec}. However, simultaneous discretization of several independent forest characteristics of a different nature substantially reduced our ability to understand the ecological meaning of model predictions, which was one of the major advantages of the Markov Chain approach \cite{pastor2005application, caswell2001, Strigul.et.all.2012, Lienard16Markov}.  

\subsection{Forests and stock markets}
In this work, we employ time series (autoregressive models and random walk) to quantify disturbance regimes and to build a predictive stochastic model of multiple disturbance classes. This type of models can overcome both major shortcomings of previous models outlined above. Autoregressive models operate with continuous space, they are analytically tractable, and they can operate with multidimensional characteristics of complex adaptive systems. Similar approaches have been successfully applied, for example, in modeling stock market fluctuations. We develop a stochastic theory of forest dynamics using an analogy to stock market theory in financial mathematics. A stock market is another complex system with random fluctuations due to multiple difficult-to-predict factors. Each individual stock has fluctuations with heavy tails. But the total stock market, measured by an index (such as Standard \& Poor 500) has long-run fluctuations (3--4 years) which follow Gaussian distribution. These fluctuations depend on various factors, such a price-earnings ratio (this measures whether the stock market is underpriced or overpriced; one can informally think of this as temperature of the stock market) and Treasury rates (long-term and short-term). These factors, in turn, can be modeled as various autoregressive models. Our main idea is to imagine that individual patches behave like stocks, and an average over a particular region is a stock market index.

\subsection{Understanding and modeling of forest patch dynamics}
In this work we propose to employ autoregressive models to understand and predict forest dynamics at the patch level. The patch-mosaic concept \cite{WuLoucks} was actively developed in the second half of the twentieth century after suggestion in \cite{Watt} that forested ecological systems can be considered a collection of patches at different successional stages. Dynamic equilibrium arises at the level of the patch mosaic rather than at the level of individual patches. The classic patch-mosaic methodology assumes that patch dynamics can be represented by changes in macroscopic variables characterizing the state of the patch as a function of time \cite{LevinPaine}. Conservation law modeling in the case of continuous time and patch state results in the reaction-advection-diffusion model \cite{LevinPaine}.

Patch dynamics concept can be applied for understanding and predicting of forest dynamics at different levels of forest organisation within the hierarchical patch dynamics paradigm \cite{WuLoucks, Strigul2012}. At the level of individual trees patch dynamics concept is traditionally called the forest gap dynamics \cite{Shugart1984, scholl2010fire, mccarthy_2001}. Individual-based forest models capture gap dynamics by simulating growth, competition and mortality of individual trees \cite{pacala1993, pacala1996, Bugmann2001, dube_at_al_2001}. Individual-based models and analytically-tractable models approximating gap dynamics \cite{Kohyama2001, Strigul2008} provide scaling from individual-level dynamics to the next level of forest hierarchical organisation, the stand-level.

In the present work we apply autoregressive models to the stand-level forest patch dynamics. At this level of forest organisation we operate with forest patches (forest stands) consisting of a large number of individual trees \cite{Strigul.et.all.2012}. Forest stand is affected not by individual-level tree dynamics, as well as by large-scale disturbances such as forest fires, drought and hurricanes \cite{hanson2000drought, mccarthy_2001}, which affect many trees in the stand at the same time. The stand-level dynamics scales up to the next hierarchical level of particular forest type or regional patch mosaic (level 3 in the hierarchical patch mosaic Matreshka model \cite{Strigul2012}).

The interplay of individual-level and stand-level changes and disturbances creates complex dynamical patterns at the stand-level. One particular source of complexity is related to a large number of intermediate level disturbances affecting only a fraction of trees in the patch \cite{Strigul.et.all.2012}. As the consequence of this system complexity a classical linear patch dynamics model does not capture patch dynamics of the US and Canadian forests \cite{Strigul.et.all.2012, Lienard2014Shade, Lienard16Markov}. This classical patch-dynamics model can be represented in continuous case by advection-reaction patch-dynamics conservation law model \cite{LevinPaine}, and in discrete case by birth and death process that can be written as a Markov Chain \cite{Strigul2012}, or as a simple forest fire model \cite{Van_Wagner_1978}. As the result, in order to capture patch dynamics of the US and Canadian forests, we need to consider more complicated models. In particular, if we discretize forest dynamics with respect to both time and state variable (biomass) we can achieve an adequate representation of forest patch dynamics within the framework of Markov chains \cite{Strigul.et.all.2012, Lienard2014Quebec, Lienard16Markov}. Markov chains provide an analytically tractable representation of forest stand dynamics, while they have a discretization error that is challenging to quantify. This work introduces an autoregressive modeling approach in application to the forest patch dynamics in Quebec. Theoretically,  our modeling approach will deliver stochastic and analytically tractable models operating with continuous state-space and -time, without discretization errors.

\subsection{Our contributions}
Here, we model dynamics of Quebec forests using a traditional AR(1) process borrowed from quantitative finance without modifications. We select the Quebec forest inventory for this proof of concept work as it is a long term dataset collected over more than 3 decades using the same field survey protocol \cite{Lienard2014Quebec}. We operate with the same biomass and basal area data derived from Quebec forest inventories in our previous publication on Markov chain modeling (data-mining protocol is available in \cite{Lienard2014Quebec, Lienard15Ecological}). For both individual patches and the Quebec region, we model logarithms of biomass or basal area as autoregressive process. The best fit, in a certain sense, turns out to be a random walk, with independent increments, which allows to quantify forest disturbance regime overall at the regional level. Regional averages are well-described by normal distribution, while individual patches are not: Fluctuations have heavy tails. This is similar to financial markets, with individual stocks having non-normal fluctuations, and stock indices (in 3--4 years or more) having normal fluctuations. To account for differing number of observations each year, we use Bayesian analysis for annual averages.

\section{Materials and Methods}

\subsection{Data mining of Quebec provincial forest inventories}

We base our research on Quebec forest inventory data 1970--2007 \texttt{www.mffp.gouv.qc.ca}. Each permanent forest inventory plot has a circular form of 400 $m^2$. The database consists of 32552 plot re-measurements at 11660 different locations. The Quebec forest inventory is designed to comprehensive describe patch mosaic of Quebec forests and plots cover the Quebec territory practically uniformly. The GIS-based map of forest inventory plots is published as Figure 1 in \cite{Lienard16Markov}. Forest inventory plots cover \emph{hardwood} and \emph{mixed forests} in the northern temperate zone (9621 and 7663 measurements, respectively) and \emph{continuous boreal forests} in the boreal zone (11969 measurements). These forest patches (forest inventory plots) are remeasured every few years often with irregular time intervals between measurements. The inventory plots were affected by natural and anthropogenic disturbances including fire and harvesting. The statistical analysis of the measurement dynamics and re-measurement intervals are published in \cite{Strigul.et.all.2012} (see Figure 2 in Appendix to \cite{Strigul.et.all.2012}). In this inventory, each patch observation  includes diameter of each tree, its species, soil moisture, and other characteristics. This is the raw data which is then converted to a more tractable data series. In particular, we are interested in biomass and basal area. Calculations of biomass and basal area were previously done according to \cite{Jenkins_et_al_2003}. The computations of biomass and basal area (as well as other characteristics, such as shade tolerance index, and biodiversity measured by Shannon entropy) is done in articles \cite{Lienard2014Quebec, Lienard15Ecological, Lienard2014Shade, Strigul.et.all.2012}. The biomass is this article refers to the plot biomass, which is the sum of biomasses of all trees computed using formulas from \cite{Jenkins_et_al_2003} (see section 3.1 in \cite{Lienard2014Quebec} for the details).  The code used for this article is available on \texttt{GitHub} repository \texttt{asarantsev/Quebec}.

\subsection{Autoregressive model for individual forest patches}

We propose a new method of modeling the biomass of an individual patch: {\it autoregressive model}, when each next year's logarithm of biomass $y(t+1)$ is a weighted sum of the previous year's logarithm of biomass $y(t)$ and a random Gaussian noise. See the primer on autoregressive models in Appendix B. We measure biomass on a logarithmic scale since it is always positive. That is,
\begin{equation}
\label{eq:AR-1}
y(t+1) = r + ay(t) + \varepsilon(t),
\end{equation}
where $r, a$ are constants, and $\varepsilon(t)$ are i.i.d. (independent identically distributed) $\mathcal N(0, \sigma^2)$ random variables. If $0 < a < 1$, this sequence $y(0), y(1), \ldots$ exhibits {\it mean reversion} to its long-term average $m = r/(1-a)$. That is, if $y(t) > m$, then $y(t+1) - m$ is likely to be smaller than $y(t) - m$, and vice versa. Examples of earlier use of such models for forest modeling include \cite{Scaling, Spatial, Silva}. They include also spatial models (incorporating distance between patches). We shall not attempt it here, instead treating every patch as effectively isolated. Building a spatial model for Quebec forest is left for future research.

Since data is collected on irregular time intervals, we apply~\eqref{eq:AR-1} multiple times to itself to get the expression of $y(t_1)$ from $y(t_0)$ if $t_0$ and $t_1$ are consecutive years for which this patch was observed. Then we try various $a$ and obtain for each $a$ the maximum likelihood estimate via linear regression. We compare these likelihoods and find the best fit for $a$. It turns out to be $a = 1$. That is, this sequence actually does not exhibit any mean reversion, but behaves like a {\it random walk}, when each next increment is independent from the past:
$$
y(t+1) = y(t) + r + \varepsilon(t),\quad \varepsilon \sim \mathcal N(0, \sigma^2)\quad \mbox{i.i.d.}
$$
The biomass itself is a {\it geometric random walk}: a process whose logarithm is random walk.  

However, the residuals for $a = 1$ do not pass the normality test. This model does not actually fit well, and we cannot find confidence intervals for $a$ using standard statistical techniques. This is due to noise in measurements of individual patches. Later in the article, we find that the average biomass over all patches exhibits more regular behavior, with normal increments. 

We perform two versions of this computation: for biomass and for basal area. For each version, we do it in two ways: (a) original logarithms of biomass/basal area; (b) with logarithm of mean biomass/basal area for this year subtracted. In both cases, the maximum likelihood estimate gives us random walk $a = 1$.

The biomass and the basal area are highly dependent, and one can plausibly use one of these metrics instead of the other.

We inherit these techniques from quantitative finance. In particular, the geometric random walk model is a classic model for the stock market movements, going back to classic research by \cite{Fama}. Mean reversion is commonly observed in financial ratios such as earnings yield or dividend yield,  \cite{JPMorgan, TimeSeries}. See also \cite{CampbellShiller, Vanguard, PD} for this research and influence of financial ratios on stock market performance. However, these techniques are less known in mathematical biology.  

\subsection{Annual averages} 
We are also interested in each year's average over all patches. We have only 38 observations for the mean value. Let $c(t)$ be the logarithm of this mean. We find that the random walk adequately describes this:
$$
c(t) = c(t-1) + \mu + \varepsilon(t),\quad \varepsilon(t) \sim \mathcal N(0, \rho^2)\quad \mbox{i.i.d.}
$$
In particular, we find that the increments $\varepsilon(t)$ indeed have normal distribution, and not heavy tails. However, we cannot simply take yearly averages for every year $t$, since they would have different precision. Reason: each year $t$ has a different number of observations. To account for this, we use Bayesian statistics. We put a prior distribution on the values of $\mu$ and $\rho^2$ (which corresponds to the lack of any existing information), and then compute the posterior distribution from the likelihood. Bayesian techniques are increasingly used in ecology, \cite{Ellison} as well as in medical statistics \cite{Medicine}, and quantitative finance \cite{BayesFinance}.

\section{Results and Discussion}

We have 32552 observations of 11660 patches of Quebec forests. Each patch is observed at most once a year, during 1970--2007. On average, each patch is observed around 3 times: $32552/11660 \approx 3$. Out of these 11660 patches, 10215 have more than one observation, which allows us to model time dynamics. Each observation consists of 4 numbers: Patch ID; year; biomass; and basal area. Various patches are observed in different years: Patch 7000406701 is observed only in  1970; patch 7000406901 is observed in 1970 and  1978; patch 8509702201 is observed in 1985, 1993, 2003; and patch 7000406902 is observed in 1970, 1978, 1985, 1997. Let $P$ be the set of observed patches. Each $p \in P$ has observations $x(t, p)$ at years $t \in T(p)$, where $T(p) \subseteq \{1970, \ldots, 2007\}$. Let $y_p(t) := \ln x(t, p)$. Yearly means are defined as:
$$
m(t) := \frac1{\#\{p \in P\mid t \in T(p)\}}\sum_{p: t \in T(p)}x(t, p),\quad t = 1970, \ldots, 2007.
$$
And we define $c(t) = \ln m(t)$. The main difficulty is that for almost all patches, a gap between consecutive observations is more than one year, and it differs from patch to patch. In particular, there are 3334 pairs of patch-year observations with the same patch and the gap 8 years, 1923 such pairs with the gap 9 years, but only 66 pairs with gap 22 years. More detailed data is in Appendix D.

\begin{center}
\begin{figure}
\label{fig:annual-means}
\centering
\subfloat[Annual Means of Biomass]{\includegraphics[width = 7cm]{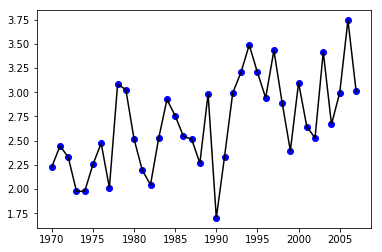}}
\subfloat[Annual Means of Basal Area]{\includegraphics[width = 7cm]{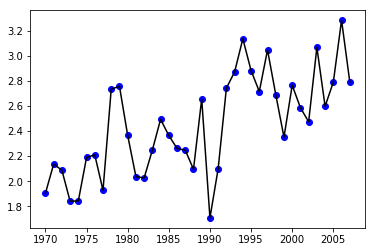}}
\caption{Annual means of biomass (measured in $10^3 kg/ha$) and basal area (measured in $m^2/ha$)}
\end{figure}
\end{center}

\subsection{Autoregressive model for individual patches}

Consider the following time series: 
\begin{equation}
\label{eq:AR}
y_p(t) = r + ay_p(t-1) + \varepsilon_p(t),\quad t = 1970, \ldots, 2007,\, p \in P,
\end{equation}
where $a$ is an AR(1) parameter, and $\varepsilon_p(t)$ are i.i.d. $\mathcal N(0, \sigma^2)$. As mentioned earlier, our main difficulty is that we do not have observations for all $t$ and $p$, only for $t \in T(p)$. Assume $t, t+u$ are subsequent time points in $T(p)$. Then 
\begin{equation}
\label{eq:AR-u}
y(t + u)  = a^uy(t) + \left(1 + a + \ldots + a^{u-1}\right)r + \sigma\sum\limits_{t=1}^{s}a^{v-1}\,\varepsilon_p(t+v).
\end{equation}
We do this both for $y_p(t)$ (raw data), $\tilde{y}_p(t) = y_p(t) - c(t)$ (centered data), and centered data without $t = 1982$ and $t = 2004$. As discussed above, these years have only very few observations, and we do not have much confidence in these values. We could write the log-likelihood of~\eqref{eq:AR-u} and apply maximum likelihood estimate using gradient descent. However, since we do not have many data points, we can use an equivalent method which is computationally inefficient but easy to implement: Fix an $a$ and run regression with respect to $r$. Then choose an $a$ such that the standard error of this regression is smallest. To properly apply this linear regression, divide~\eqref{eq:AR-u} by a constant to make the standard error in error terms in~\eqref{eq:AR-u} the same for all $u$:
\begin{align}
\label{eq:auxillary-regression}
\begin{split}
C(a, u)\left(y(t+u) - a^uy(t)\right) & = D(a, u)\cdot r + \delta_p(t+u),\quad \delta_p(t+u) \sim \mathcal N(0, \sigma^2)\quad \mbox{i.i.d.}\\
C(a, u) & := \left[1 + a^2 + \ldots + a^{2(u-1)}\right]^{-1/2},\\
D(a, u) & := C(a, u)\left[1 + a + \ldots + a^{u-1}\right].
\end{split}
\end{align}
For every $a \in (-2, 2)$ which is a multiple of $0.01$, fit a simple linear regression (without intercept) to find $r$ and the standard error $\sigma$. Appendix A explains why minimizing standard error given normalized residual variance is equivalent to maximizing likelihood. We do this analysis three times: for original values, centered values, and centered values with years 1982 and 2004 removed. We repeat this for biomass and basal area. For all six cases, the parameter $a = 1$ minimizes the standard error (thus maximizing the likelihood). Thus, the dynamics in~\eqref{eq:AR} is given by $y_p(t) = r + y_p(t-1) + \varepsilon(t)$, which is a simple random walk with Gaussian increments.

\begin{center}
\begin{figure}
\centering
\subfloat[Standard error $\sigma$ vs $a$]{\includegraphics[width = 6cm]{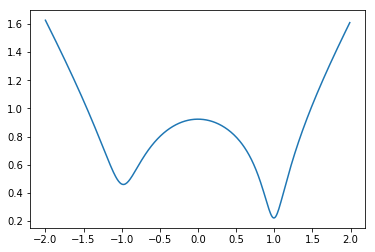}}
\subfloat[QQ plot for residuals]{\includegraphics[width = 6cm]{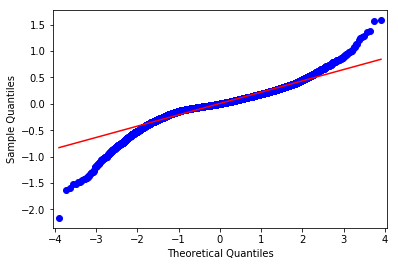}}
\caption{Non-centered biomass data for individual patches, with $\sigma$ measured in $10^3kg/ha$}
\end{figure}
\end{center}
Assuming that the model is, in fact, a random walk, let us make quantile-quantile (QQ) plots for residuals in~\eqref{eq:auxillary-regression} (we have simple linear regression {\it without} intercept):
$$
u^{-1/2}(y(t+u) - y(t)) = r\sqrt{u} + \sigma\varepsilon_p(t).
$$
These QQ plots of residuals in this regression~\eqref{eq:auxillary-regression} are not normal. Thus more analysis is needed. 

For non-centered biomass, the minimal standard error is achieved when $a = 1$, see Figure 2 (A). This corresponds to a random walk model, and we get $\sigma = 0.223$ and $r = 0.0378$.  However, the residuals are not normal: See them on the QQ plot in Figure 2 (B). Then we repeat the computation above for centered values: $\tilde{y}_p(t) = y_p(t) - z(t)$ instead of $y_p(t)$. Similarly, the standard error is minimized for $a = 1$. For this value, $r = 0.0569$, $\sigma = 0.224$. The QQ plot of residuals is still not normal. For centered values with years 1982 and 2004 removed, the standard error once again is minimized for $a = 1$, with $\sigma = 0.224$ and $r = 0.00570$. The QQ plot of residuals is still not normal. We omit the standard error graph and the QQ plot for the last two cases: centered data with all years, and centered data without 1982 and 2004, since these plots are very similar to their counterparts for original (non-centered) data. 

Modeling basal area data as in~\eqref{eq:AR}, and computing standard error of the regression~\eqref{eq:auxillary-regression}, we again get $a = 1$, $r = 0.0337$, and $\sigma = 0.207$. Again, the best-fitting model among AR(1) according to the maximum likelihood estimation is the random walk. If we center the basal area data, and consider all years, then again, the standard error is minimized for $a = 1$, with $r = 0.00340$ and $\sigma = 0.197$. Centering the basal area data and removing years 1980 and 2004, we get:  $a = 1$, $r = 0.00341$, $\sigma = 0.197$. In all these cases, similarly to the case of biomass, the QQ plots of residuals for basal area are not normal, with both tails fat. We do not provide pictures of QQ plots, since they are very similar to that for biomass.   
Correlation between biomass and basal area: Take all patches $p$ and corresponding years $t$ in $T(p)$ Denote by $y'_p(t)$ the logarithm of biomass, and by $y''_p(t)$ the logarithm of basal area for patch $p$ and year $t$. If $T(p) = \{t_0, t_1, t_2, \ldots\}$ has more than one year, order them in increasing order: $t_0 < t_1 < t_2 < \ldots$ Compute correlation coefficient between 
$$
y'_p(t_k) - y'_p(t_{k-1})\quad \mbox{and}\quad 
y''_p(t_k) - y''_p(t_{k-1}).
$$
It is equal to 0.983. With years 1982 and 2004 removed, this number does not change (in the first four decimal digits). Thus biomass and basal area for individual patches are very correlated. For practical purposes, this means we can use either measure as a size of patch. 


\begin{center}
    \begin{figure}
    \centering
    \subfloat[Biomass 2007]{\includegraphics[width = 7.5cm]{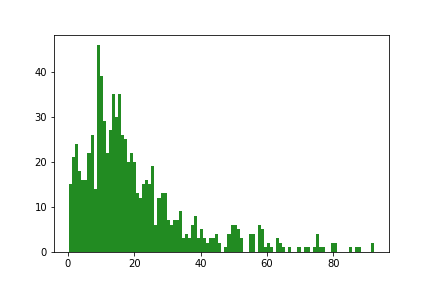}}
\subfloat[Basal area 2007]{\includegraphics[width = 7.5cm]{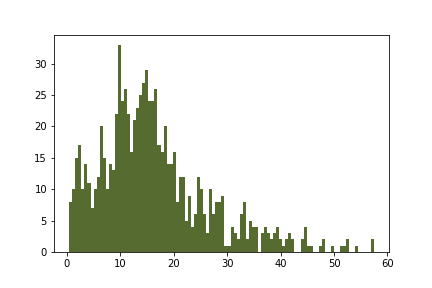}}
\caption{Histograms for biomass (measured in $10^3kg/ha$) and basal area (measured in $m^2/ha$) in 2007}
\end{figure}
\end{center}

\begin{center}
    \begin{figure}
    \centering
\subfloat[Biomass 2019]{\includegraphics[width = 7cm]{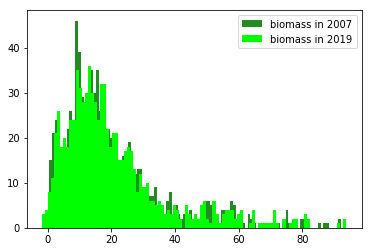}}
\subfloat[Basal area 2019]{\includegraphics[width = 7cm]{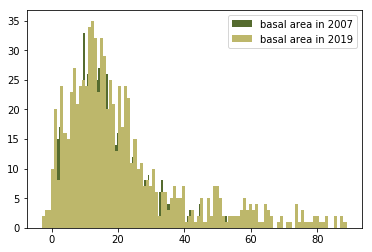}}
\caption{Histograms for biomass (measured in $10^3kg/ha$) and basal area (measured in $m^2/ha$) in 2019 superimposed upon 2007}
    \end{figure}
    \end{center}

\subsection{Yearly averages, frequentist analysis}

We model the logarithm of yearly average using random walk:
\begin{equation}
\label{eq:increments}
\triangle c(t) := c(t+1) - c(t) \sim \mathcal N(\mu, \rho^2)\quad \mbox{i.i.d}
\end{equation}
However, years 1982 and 2004 have only a few observations, see the table from Appendix D. Thus we do not have much confidence in these results. Thus we do the second test: Remove these years, and consider increments~\eqref{eq:increments} for 
$t = 1970, \ldots, 1981, 1984, \ldots, 2003, 2006, 2007$. Then we test normality of increments using the QQ plot and Shapiro-Wilk test. We confirm that, indeed, increments are i.i.d. normal. Rewrite the model~\eqref{eq:increments} as the random walk:
\begin{equation}
\label{eq:random-walk-average}
c(t) = c(0) + \xi_1 + \ldots + \xi_t,\quad \xi_i \sim \mathcal N(\mu, \rho^2)\quad \mbox{i.i.d.}
\end{equation}
Or, equivalently, we can rewrite~\eqref{eq:random-walk-average}  in the original scale, instead of the logarithmic scale:
\begin{equation}
\label{eq:GBM-average}
m(t) = m(0)\exp\left(\xi_1 + \ldots + \xi_t\right),\ t = 0, 1, 2, \ldots
\end{equation}
From~\eqref{eq:GBM-average}, we can compute the mean and variance of $m(t)$:  
\begin{align*}
\mathrm{E}[m(t)] = m(0) \cdot \exp \big(t(\mu+\sigma^2/2)\big); \quad 
\mathrm{Var}[m(t)] = m^2(0) \cdot \exp (2\mu t+\sigma^2 t ) \cdot \big(\exp (\sigma^2 t )-1\big).
\end{align*}

For the biomass, we get: $\mu = 0.021$ and $\rho = 0.512$. These increments passes Shapiro-Wilk normality test with $p = 0.80$. With removed two years 1982 and 2004, the estimates will not change much:  $\mu = -0.011$ and $\rho = 0.507$. This still passes Shapiro-Wilk normality test with $p = 0.60$. Repeating this analysis for basal area instead of biomass, we get: $\mu = 0.0240, \rho = 0.367$ for all years, and $\mu = 0.00455, \rho = 0.370$ for years without 1982 and 2004. Shapiro-Wilk test gives us $p = 0.89$ for all years, and $p = 0.58$ for years without 1982 and 2004. The QQ plots in Figure 5 show that, indeed, the residuals are close to normal. 

In Figure 3, we plot histograms of the biomass and basal area for an individual patch in 2007. In Figure 4, we simulate biomass and basal area as in~\eqref{eq:random-walk-average} until 2019, starting from a patch randomly selected among observed patches in 2007. We superimpose this histogram upon the one for 2007. 

Taking increments of logarithms of yearly means for biomass and basal area (37 data points), we get correlation 0.977. For years 1982 and 2004 removed, we get correlation 0.980. Previously, we got very high correlation between increments of logarithms for individual patches, thus we conclude that biomass and basal area are the same for practical purposes, as measures of size. Now we see that the same is true for yearly averages. 

\begin{center}
\begin{figure}
\centering
\subfloat[All Years]{\includegraphics[width = 6cm]{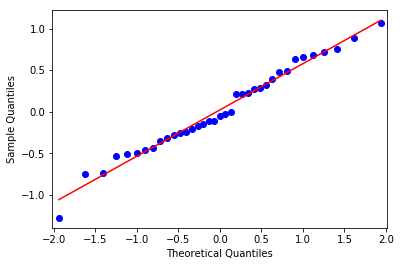}}
\subfloat[Without 1982 and 2004]{\includegraphics[width = 6cm]{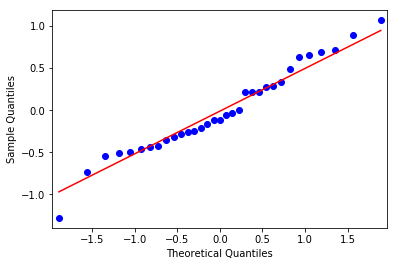}}
\caption{QQ plot for biomass yearly average logarithm increments}
\end{figure}
\end{center}

\subsection{Yearly averages, Bayesian analysis}

As mentioned earlier, the analysis in the previous section is deficient: The means have different precision for different years, because the quantity of patches observed differs from year to year. Thus we apply Bayesian statistics in this section. 

A primer on Bayesian statistics for normal distribution can be found in Appendix C. Let for each fixed year $t$  the logarithms of observations are $x_1(t), \ldots, x_n(t)$. For these values we performed analysis above: The posterior mean $\mu(t)$ and the posterior variance $v(t)$ were generated. The simulation of $\mu(t)$ and $v(t)$ was performed $N = 1000$ times. In Figure 6, we have histograms of 1000 simulations for $\mu(t)$ and $\sigma(t)$ for $t = 0$ (year 1970), for both biomass and basal area. Hence we obtain 38 sequences of $N$ numbers: $\mu_1(t), \ldots, \mu_N(t)$, $t = 1970, \ldots, 2007$. The average growth rate based on simulated results is 
\begin{equation}
\label{eq:G}
\hat{g} = \frac1{NT}\sum\limits_{i=1}^{N}\big[\mu_i(T) - \mu_i(0)\big].
\end{equation}
The mean increments are: $\triangle\mu_i(t) = \mu_i(t) - \mu_i(t-1),\quad t = 1, \ldots, T,\quad i = 1, \ldots, N$. Assuming these are $\mathcal N(g, \sigma^2)$ i.i.d. we estimate $g$ and $\sigma^2$ as $\hat{g}$ in~\eqref{eq:G} and $\hat{\sigma}^2$:
$$
\hat{\sigma}^2 = \frac1{NT}\sum\limits_{i=1}^N\sum\limits_{t=1}^T(\triangle\mu_i(t) - \hat{g})^2.
$$
We compute the point estimates of $g$ and $\sigma^2$ for biomass and basal area:
$$
\hat{g}_{bio} \approx 0.023,\quad \hat{\sigma}_{bio}^2 \approx 0.214;\quad \hat{g}_{BA} \approx 0.023, \quad \hat{\sigma}_{BA}^2 \approx 0.134.
$$
Thus the growth rate of forest, measured by the biomass or basal area (on the logarithmic scale), is $2.3\%$ per year. These numbers are close to the ones from frequentist analysis from the previous subsection: $2.1\%$ with years 1982 and 2004. However, with removed years 1982 and 2004, this estimate changes to $-1.1\%$.
For basal area, we have a similar comparison with the previous subsection: $2.4\%$ with 1982 and 2004, $0.46\%$ without these years. As discussed earlier, we view Bayesian analysis as the more statistically sound.
Thus $2.1\%$ growth per year seems more reasonable. 

From year to year, however, there are a lot of fluctuations, or, to use a stock market term, volatility: The standard deviation for yearly fluctuations is estimated as $\sqrt{0.214} = 0.46$, that is, $46\%$ per year for the biomass, and $\sqrt{0.134} = 0.37$, that is, $37\%$ per year for the basal area. Similarly to the mean estimates, we view these as more scientifically sound that the ones from frequentist analysis from the previous subsection.

\begin{center}
\begin{figure}
\label{fig:mean-biomass-BA}
\centering
{\includegraphics[width = 7cm]{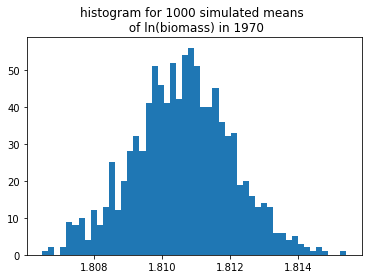}}
{\includegraphics[width = 7cm]{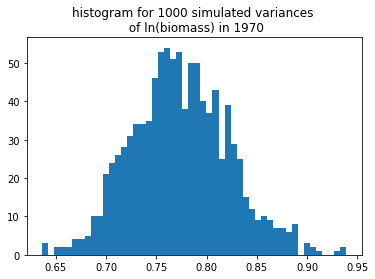}}
\\
{\includegraphics[width = 7cm]{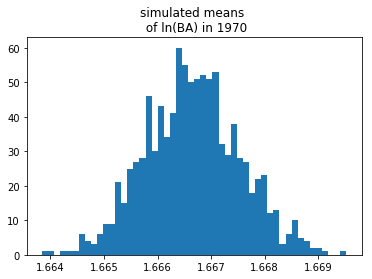}}
{\includegraphics[width = 7cm]{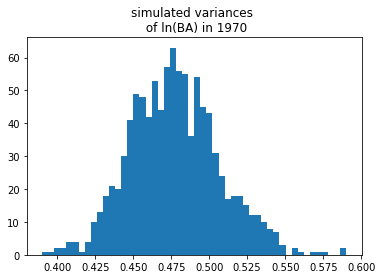}}
\caption{Histograms of $N = 1000$ simulations for means and variances of the logarithm for biomass and basal area, 1970. The biomass is measured in $10^3kg/ha$, and the basal area is measured in $m^2/ha$.}
\end{figure}
\end{center}

\section{General discussion}

\subsection{Towards autoregressive theory of forest dynamics}
We have applied AR(1) autoregressive model to patch/stand dynamics of Quebec forests. Overall, we followed major steps of application of autoregressive model in financial engineering considering stand biomass and basal area as variables similar to a stock market index. However, forest and stock market are different complex systems despite the observed similarities. We consider this work as the first step, and the further discussion is required: it is hard to expect that the forest dynamics modeling will simply mirror stock market theory.

One interesting result reported in the Section 2.2 is that $a = 1$ and the residuals do not pass the normality test. In simple terms, it means that each forest patch is highly volatile, and its dynamics cannot be well described by the normal distribution. Still, among all AR(1), random walk has the maximum likelihood. This suggests for possible use random-walk models with tails heavier than normal. This dichotomy of average patch vs individual patch reminds us of a stock market dynamics, where each individual stock is highly volatile, with non-Gaussian fluctuations \cite{Heavy}, but a stock market overall (measured by Standard \& Poor 500 or another stock market index) in the long run follows geometric random walk with normal increments (if the time step is large enough, say 3-4 years or more).

 From the perspective of random processes $a = 1$ means that we are dealing with the divergent AR(1) process, traditionally called the random walk, which does not have a stationary distribution. The AR(1) converges to its stationary distribution when $-1 < a < 1$. This means that biomass and basal area at every step will drift away like in both positive and negative directions. We can search for mechanistic meaning of this result in both fundamental patterns of forested ecosystem dynamics and in general modeling assumptions related to the autoregressive modeling approach and the patch dynamics modeling framework.

Patch dynamics modeling framework assumes that each patch, a forest stand in our case, has the same underlying transition probabilities related to internal changes such as tree growth as well as with respect to natural disturbances \cite{Strigul.et.all.2012}. These are typical background assumptions of practically all forest models, however the validity of these assumptions should be evaluated. In particular, in our application we consider the combined data set consisting of hardwood and mixed temperate forest stands as well as continuous boreal forests. In addition, some of the stands even in the same forest types are located in  different climatic conditions and might be affected by various disturbance regimes. The overall robustness of modeling predictions with respect to this type of data variability is typically known for traditional models. We have previously investigated effects of these assumptions in the Markov chain modeling framework \cite{Strigul.et.all.2012, Lienard2014Quebec, Lienard16Markov}. Similarly, individual-based models are often applied for forest simulations in different regions with the same growth/mortality characteristics of particular tree species. Autoregressive modeling is a new approach in forest modeling, and it is not yet known how robust our predictions with respect to variability of forest patch mosaic. We can hypothesise that outcomes reported in the Section 2.2 ($a = 0$ and the normality test for residuals) might be related to the variability in forest inventories.

Another common underlying assumption in forest modeling is that the random process is time homogeneous or stationary \cite{Strigul.et.all.2012}. It is often assumed that climatic and disturbance regime changes occur at the slower scale and can be ignore for the short- and intermediate-term modeling \cite{Lienard2014Quebec}. Time inhomogeneous Markov chains allow to relax this assumption and take into account both climate driven changes in tree growth and frequency of environmental disturbances \cite{Lienard16Markov}. Similarly to inhomogeneous Markov chain theory, autoregressive models can be generalized to include directional changes in disturbance dynamics. This leads us to autoregressive integrated moving average (ARIMA) models, which can capture non-stationary (time inhomogenous) dynamics. In particular, we previously detected some non-stationary effects related to the disturbance regimes recorded in the Quebec forest inventory, such as small decrease of the total disturbance rate in boreal and hardwood forests (see Table 1 in \cite{Lienard16Markov}). However, the relatively small number of the measurements in the data set did not allow us to report some definite trend. Nevertheless, we anticipate that the pameterization and validation of ARIMA models can be an important step towards time series theory of forest patch dynamics.

Time series analysis can also be applied to forest dynamics at different hierarchical levels. In a certain way autoregressive models are continuous time and state counterpart of discrete Markov chain models. Markov chains were applied at the individual tree and species levels \cite{waggoner_stephens_1970,stephens_waggoner_1980}, stand-level \cite{usher1979markovian}, and landscape-level \cite{LogofetLesnaya2000}. The autoregressive and ARIMA models can be also employed to capture similar dynamics. In addition, time series models might be used in approximation of the forest gap dynamics and individual-based models.

\subsection{Future research}

 That the random walk with Gaussian increments performed better than any other AR(1) suggests that there is not much dependence of annual change of the biomass on the current biomass. However, testing this statement requires further research.
 
Time-series analysis is a novel modeling approach for forest modeling at the large scale. We wish to compare  this approach with state-of-art mathematical models developed on other principles, in particular with Markov chains \cite{Strigul.et.all.2012, Lienard2014Quebec, Lienard15Ecological, Lienard2014Shade} and individual-based models \cite{Strigul2012, Lienard16IBM}, developed using the same forest inventory data sets. We also need to understand the merits of discrete vs continuous space. This is critical for practical applications of forest models for natural resource management, and risk assessment of forest vulnerability to changes in environmental disturbance regimes.

We can generalize time series approach to model forest dynamics and climate at the regional scale. Such generalization in financial engineering resulted in the development of vector autoregressive integrated moving average (VARIMA) models, which are particularly suited for multidimensional forest modeling under the dynamic disturbance regimes. This can result in an enhanced tolerance-disturbance model build upon our past analyses of tolerance patterns in North American forests, including observed shade tolerance patterns \cite{Lienard2014Shade}, associations between shade tolerance and soil moisture \cite{Lienard15Ecological}, and relative trade-offs between shade, drought, and water-logging tolerance at the continental scale \cite{lienard2015DroughtTDM_fia}. 

We can extend our research to the USA Forest Inventories: Analysis of current disturbance regimes across US ecoregions, and how disturbance regimes are connected with other forest macroscopic characteristics. In particular, forest tolerance is a key ecological driver that controls the response of forested ecosystems to climate change-related disturbances such as drought- and fires \cite{lienard2015DroughtTDM_fia}. We can link climatic models and forest tolerances \cite{Lienard2014Shade, Lienard15Ecological, lienard2015DroughtTDM_fia}. Spatially-explicit information from the USDA Forest Service Forest Inventory and Analysis Program (FIA), Canadian Forest Inventory, and climate data sets and models (WorldClim and PRISM), can be integrated to quantify the effect climate variables in North America have on forest tolerances and disturbances via recently proposed methodology \cite{lienard2015DroughtTDM_fia}.

\section{Conclusion}

Individual patches have biomass and basal area whose dynamics (on logarithmic scale) are not well described by classic autoregressive model  
$y(t+1) = ay(t) + r + \varepsilon(t)$ with normal noise terms $\varepsilon(t)$, because the tails of these noise terms turn out to be heavier than normal. However, among these AR(1) models, the best fit (according to the maximum likelihood) is random walk, with $a = 1$, and increments $y(t+1) - y(t)$ independent of the past $y(s), s \le t$. Thus one can try a heavy-tailed random walk, in which increments have tails heavier than Gaussian. This topic is left for future research. 

In contrast, yearly means (on the log scale) are well described by random walk with normal increments. Bayesian analysis accounts for the fact that different years have different number of observations. We get growth rate (measured on the log scale) $2.3\%$ per year, with standard deviation $46\%$ for biomass, and $37\%$ for basal area. Thus the forest grows on average in the long run, but from year to year there is a lot of volatility. 

An important part of forest ecological modeling is to quantify disturbances from fires, droughts, etc. These events quickly destroy significant parts of the forest. An analogy for the stock market would be a crash, as in 2001 or 2008. Indeed, for the stock market, 1-year fluctuations are not adequately described by the normal distribution (only 3--4 years or more are). But for the forest as a whole (as opposed to particular patches), annual fluctuations are normal. We do not need to introduce separate distributions for modeling disturbances. Since we assume the prior is Jeffrey's non-informative and the model is Gaussian, we do not need to do any Monte Carlo or Metropolis-Hastings computations: There are explicit formulas for the posterior distribution of parameters.

\section*{Acknowledgments} 

We thank Dr. Adam Erickson for help with proofreading. Initial data-mining of Quebec forest inventories was done by Dr. Jean Lienard for the Markov chain modeling and published in \cite{Lienard2014Quebec}. We also acknowledge welcoming environment for applied quantitative research in our departments. A.S. is grateful to Dr. Anna Panorska and Dr. Tomasz Kozubowski for mentorship and support. This work was partially supported by a grant from the Simons Foundation (\# 283770 to N.S.)

\section{Appendix}
\subsection{Maximal likelihood and minimal standard error}

Take a family of linear regressions depending on the parameter $a$, with $d$ factors and the intercept:
\begin{equation}
\label{eq:example-app}
y_i = c_0(a) + c_1(a)x_{i1} + c_2(a)x_{i2} + \ldots + c_d(a)x_{id} + \varepsilon_i(a),\quad \varepsilon_i(a) \sim \mathcal N(0, \sigma^2)\quad \mbox{i.i.d.}
\end{equation}
Solve for $a$ by maximum likelihood estimation. The Gaussian density for $\mathcal N(0, \sigma^2)$ is given by \begin{equation}
\label{eq:Gaussian}
p(x\mid \sigma) = \frac1{\sqrt{2\pi}\sigma}\exp\left[-
\frac{x^2}{2\sigma^2}\right].
\end{equation}
The log likelihood of~\eqref{eq:example-app} is derived from the Gaussian density~\eqref{eq:Gaussian} and is given by 
$$
\ell = -\frac d2\ln(2\pi) - d\ln\sigma - \frac1{2\sigma^2}\sum\limits_{i=1}^d(y_i - c_0(a) - c_1(a)x_{i1} - \ldots - c_d(a)x_{id})^2.
$$
But the standard error of the regression~\eqref{eq:example-app} is estimated as
$$
s^2(a) = \frac1{n-d-1}\sum\limits_{i=1}^d(y_i - c_0(a) - c_1(a)x_{i1} - \ldots - c_d(a)x_{id})^2.
$$
Thus we can express
\begin{equation}
\label{eq:connection}
\ell = -\frac d2\ln(2\pi) - d\ln\sigma - \frac{s^2(a)(n-d-1)}{2\sigma^2}.
\end{equation}
We used crucially here that the residuals in~\eqref{eq:example-app} are normalized so that they have the same variance $\sigma^2$. To maximize $\ell$ from~\eqref{eq:connection}, we need to first minimize the standard error $s^2(a)$ by computing it for each $a$ and then choosing an appropriate $a$; then to choose the $\sigma$ which maximizes~\eqref{eq:connection} for given $s^2(a)$, which turns out to be $\sigma^2 = s^2(a)$.

\subsection{Background on autoregressive models and random walk}

Consider a time series of random variables $(x_0, x_1, x_2, \ldots)$:
$$
x_n = r + ax_{n-1} + \varepsilon_n,\quad \varepsilon_n \sim \mathcal N(0, \sigma^2)\quad \mbox{i.i.d.}
$$
This is called an {\it autoregressive process of order 1} or AR(1): Regression of this sequence onto itself with a one-step time lag. It has the following properties. For $-1 < a < 1$, this sequence converges weakly to the stationary distribution: $\mathcal N(m, \rho^2)$, with  
$$
m = \frac{r}{1 - a},\quad \rho^2 = \frac{\sigma^2}{1 - a^2}.
$$
This means that for every interval $[c, d]$, 
$$
\mathbb P(c \le x_n \le d) \to \frac1{\sqrt{2\pi}\rho}\int_c^d\exp\left(-\frac{-(z-m)^2}{2\rho^2}\right)\,\mathrm{d}z,\quad n \to \infty.
$$
In addition, mean and variance of $x_n$ converge to that of the limiting distribution: $\mathrm{E}\left[x_n\right] \to m$, $\Var\left[x_n\right] \to \rho^2$. Thus this time series exhibits {\it mean reversion:} If $x_n > m$ then $x_{n+1}$ is more likely to decrease compared to $x_n$ than to increase. For $a = 1$, this is {\it random walk:} Increments $x_{n+1} - x_n$ are independent for different $n$. This sequence does not have a limit as $n \to \infty$. The expectation $\mathrm{E}\left[x_n\right] = \mathrm{E}\left[x_0\right]$ is constant. But the variance $\Var[x_n] = \Var[x_0] + n\sigma^2$. 

\subsection{Background on Bayesian inference}

Assume we have a sample $x_1, \ldots, x_N \sim \mathcal N(m, \sigma^2)$. Denote the variance by $\sigma^2 = v$. Random variables $x_1, \ldots, x_n$ are independent, and $\mathcal N(m, v)$ has density
$$
\ell(x_i\mid m, v) = (2\pi v)^{-1/2}\exp\left(-\frac{(x_i-m)^2}{2v}\right).
$$
We set a {\it non-informative prior} on $(m, v)$, which means we do not have any existing information about these parameters: $\pi(m, v) \propto v^{-1}$. This is an infinite measure: 
$$
\int_{-\infty}^{+\infty}\int_0^{\infty}\frac1v\,\mathrm{d}v\,\mathrm{d}m = +\infty.
$$
Thus we cannot normalize it (divide it by a constant) to make it a probability measure. But we can still apply Bayesian statistics to this measure. We choose this form of measure because we can get explicit posterior. Bayesian inference works as follows.
The {\it likelihood}, that is, density of $x_1, \ldots, x_n$ given $m, v$ is
\begin{align}
\label{eq:likelihood}
\begin{split}
L(x_1, \ldots, x_n\mid m, v) &= \ell(x_1\mid m, v)\cdot\ldots\cdot\ell(x_n\mid m, v) \\
& = (2\pi v)^{-n/2}\exp\Bigl(-\frac1{2v}\sum\limits_{i=1}^n(x_i - m)^2\Bigr);\\
p(m, v\mid x_1, \ldots, x_n) &\propto \pi(m, v)\cdot L(x_1, \ldots, x_n\mid m, v).
\end{split}
\end{align}
Unlike the prior, the posterior is a finite measure. We can normalize it by computing its integral and dividing it by this integral.
After computation, we get:
\begin{align}
\label{eq:explicit}
\begin{split}
p(m, v\mid x_1, \ldots, x_n) &= \frac{S^{n/2}n^{1/2}}{\Gamma(n/2)(2\pi)^{1/2}}v^{-(n+3)/2}\exp\biggl(-\frac Sv\biggr)\exp\biggl(-\frac{n(m - \overline{x})^2}{2v}\biggr)\\ \mbox{with}\quad \overline{x} & := \frac1n\sum_{i=1}^nx_i\quad \mbox{and} \quad S := \frac1n\sum\limits_{i=1}^n(x_i - \overline{x})^2.
\end{split}
\end{align}
Recall that Gamma distribution with shape $\alpha$ and scale $\beta$ has density and expectation:
\begin{align}
\label{eq:gamma}
\begin{split}
&\mbox{density}\quad f(z; \alpha, \beta) = \frac{\beta ^{\alpha}}{\Gamma(\alpha)}z^{\alpha - 1}e^{-\beta z},\quad z > 0;\\
&\mbox{mean}\quad \int_0^{\infty}zf(z; \alpha, \beta)\,\mathrm{d}z = \frac{\beta}{\alpha}.
\end{split}
\end{align}
Using~\eqref{eq:gamma}, we can rewrite~\eqref{eq:explicit} as follows:
$$
v^{-1} \sim \Gamma\left(\frac{n-1}{2}, \frac {nS}{2}\right),\quad m\mid v \sim \mathcal N(\overline{x}, v/n).
$$
That is, $v^{-1}$ has marginal Gamma distribution with shape $n/2$ and expectation $1/S$; and $v$ has inverse Gamma distribution with shape $n/2$. The conditional distribution of $m$ given $v$ is normal. The unconditional (marginal) distribution of $m$ is Student ($t$-distribution). A Student distribution has heavier tails than a normal distribution, which implies more uncertainty, resulting from our Bayesian estimation framework.

\newpage

\subsection{Empirical data}

Empirical means $\overline{x}(t)$ and variances $S(t)$ of biomass and basal area logarithms, for each year

\bigskip

{\footnotesize
\begin{tabular}
{| c | c | c | c | c | c |}
\hline
Year & Number  & Yearly Mean  & Yearly Variance  & Yearly Mean  & Yearly Variance \\
 & of & of Biomass & of Biomass & of Basal Area & of Basal Area \\
 & Observations & Logarithm  & Logarithm & Logarithm & Logarithm \\
\hline
1970  &  522  &  1.81  &  0.77 &  1.67  &  0.48 \\
1971  &  1216  &  2.05  &  0.8 &  1.9  &  0.5 \\
1972  &  1286  &  1.97  &  0.67 &  1.87  &  0.43 \\
1973  &  335  &  1.66  &  0.59 &  1.64  &  0.39 \\
1974  &  304  &  1.68  &  0.55 &  1.66  &  0.36 \\
1975  &  902  &  1.97  &  0.66 &  1.96  &  0.54 \\
1976  &  1883  &  2.17  &  0.61 &  2.02  &  0.4 \\
1977  &  422  &  1.82  &  0.37 &  1.81  &  0.25 \\
1978  &  1319  &  2.77  &  0.7 &  2.53  &  0.48 \\
1979  &  1339  &  2.68  &  0.77 &  2.52  &  0.54 \\
1980  &  1047  &  2.2  &  0.7 &  2.13  &  0.52 \\
1981  &  396  &  1.86  &  0.7 &  1.8  &  0.51 \\
1982  &  8  &  1.98  &  0.12 &  1.97  &  0.12 \\
1983  &  98  &  2.23  &  0.66 &  2.06  &  0.43 \\
1984  &  358  &  2.65  &  0.57 &  2.32  &  0.36 \\
1985  &  629  &  2.38  &  0.75 &  2.15  &  0.47 \\
1986  &  665  &  2.24  &  0.62 &  2.07  &  0.4 \\
1987  &  732  &  2.22  &  0.62 &  2.05  &  0.42 \\
1988  &  604  &  1.98  &  0.56 &  1.92  &  0.34 \\
1989  &  1597  &  2.49  &  0.95 &  2.38  &  0.58 \\
1990  &  723  &  1.46  &  0.47 &  1.54  &  0.34 \\
1991  &  581  &  2.02  &  0.61 &  1.92  &  0.38 \\
1992  &  1782  &  2.67  &  0.68 &  2.54  &  0.46 \\
1993  &  865  &  2.88  &  0.77 &  2.65  &  0.55 \\
1994  &  647  &  3.21  &  0.67 &  2.93  &  0.48 \\
1995  &  625  &  2.85  &  0.77 &  2.66  &  0.52 \\
1996  &  858  &  2.57  &  0.86 &  2.43  &  0.68 \\
1997  &  2247  &  3.08  &  0.81 &  2.82  &  0.57 \\
1998  &  977  &  2.6  &  0.68 &  2.49  &  0.49\\
1999  &  905  &  2.11  &  0.67 &  2.13  &  0.54 \\
2000  &  101  &  2.62  &  1.0 &  2.46  &  0.74 \\
2001  &  756  &  2.47  &  0.38 &  2.45  &  0.3 \\
2002  &  309  &  2.32  &  0.47 &  2.32  &  0.39 \\
2003  &  3414  &  3.08  &  0.85 &  2.83  &  0.61 \\
2004  &  19  &  2.55  &  0.29 &  2.51  &  0.23 \\
2005  &  641  &  2.71  &  0.73 &  2.57  &  0.57 \\
2006  &  599  &  3.42  &  0.81 &  3.08  &  0.53 \\
2007  &  841  &  2.67  &  0.81 &  2.55  &  0.62 \\
\hline
\end{tabular}
}

\bigskip

Quantity of observation patch-year pairs with given time gap

\bigskip

{\footnotesize
\begin{tabular}
{|c|c|c|c|c|c|c|c|c|c|c|c|c|}
\hline
Year Gap & 3 & 4 & 5 & 6 & 7 & 8 & 9 & 10 & 11 & 12 & 13 & 14\\
\hline
Quantity & 1 & 2 & 65 & 915 & 1381 & 3334 & 1923 & 2214 & 2543 & 2677 & 1972 & 694\\
    \hline
Year Gap & 15 & 16 & 17 & 18 & 19 & 20 & 21 & 22 & 23 & 24 & 25 & 26\\
    \hline
Quantity & 922 & 533 & 391 & 569 & 390 & 123 & 8 & 66 & 8 & 17 & 22 & 22\\
\hline
Year Gap & 27 & 28 & 29 & 30 & 31 & 32 & 33 & 34 & 35 & 36 & 37 & 38\\
    \hline
Quantity & 4 & 17 & 14 & 9 & 6 & 5 & 12 & 1 & 0 & 5 & 2 & 0\\
\hline
\end{tabular}
}

\bibliographystyle{plain}

\end{document}